\documentstyle[12pt]{article}
\let\section=\subsection  \let\subsection=\subsubsection

\def\be{\begin{equation}}
\def\ee{\end{equation}}

\begin{document}
\begin{center}
{\large \bf HOT NUCLEONS IN CHIRAL SOLITON MODELS}\\[5mm]
H.~WALLISER \\[5mm]
{\small \it Fachbereich Physik, Universit\"{a}t-GH-Siegen,\\ 
D-57068 Siegen, Germany \\[8mm]} 
\end{center}

\begin{abstract}\noindent
Chiral lagrangians as effective field theories of QCD are
most suitable for the study of nucleons in a hot pion gas
because they contain pions and also baryons as solitons of the same
action. The semiclassical treatment of the soliton solutions
must be augmented
by pionic fluctuations which requires renormalisation to 1-loop, 
and finite temperatures do not introduce new ultraviolet divergencies
and may easily be considered.
Alternatively, a renormalisation scheme based on the renormalisation
group equation at finite temperature
comprises and extends the rigorous
results of chiral perturbation theory and renders the low energy
constants temperature-dependent which allows the construction of
temperature-dependent solitons below the critical temperature.
The temperature-dependence of the baryon energy and the pion-nucleon
coupling is studied. There is no simple scaling law for the 
temperature-dependence of these quantities.
\end{abstract} 

\section{Introduction}
It is generally believed that with increasing temperature hadronic
matter undergoes a phase transition to a quark gluon plasma which 
hopefully might be produced in heavy ion collisions. Already below the
critical temperature we expect the baryon properties to change, of
particular interest are variations in the nucleon mass $M$ and the
pion-nucleon coupling constant $g_{\pi NN}$, which determine the
behaviour of hot nucleonic matter.

Eletsky and Kogan \cite{ek94} use ChPT (chiral perturbation theory) to
show that the temperature-dependence of the axial coupling constant $g_A$
turns out to be the same as for the pion decay constant $f_\pi$. With
the assumption that the baryon mass is temperature-independent and
with the Goldberger-Treiman relation $f_\pi g_{\pi NN} = M g_A$
they obtain
\be
\label{ChPT}
M \sim 1 , \qquad g_A \sim f_{\pi} , \qquad g_{\pi NN}
\sim 1  
\ee
a pion-nucleon coupling which remains essentially
unchanged. 

Bernard and Meissner \cite{bm91}
use a chiral soliton model with explicit vector
mesons together with a temperature-dependent $f_{\pi}$ lent from ChPT
\cite{gl87} or from the NJL model \cite{bmz87}. 
Qualitatively their results may be understood
by the simple scaling
\be
\label{scaling}
M \sim f_{\pi} , \qquad g_A \sim  g_{\pi NN} \sim 1 \, ,  
\ee
which in tree approximation becomes exact in a pure Skyrme model 
(without pion mass term)
and which is at variance with (\ref{ChPT}) although with the same
result of a temperature-independent pion-nucleon coupling. 

The simple relations (\ref{scaling}) applied to the density dependence of these
quantities have nowadays come to be known as Brown-Rho scaling \cite{br91}.
We do not want to discuss here what happens if the density is varied,
but concerning the temperature there are definitely no reasons that these
relations should hold. Namely
the temperature-dependence is carried by the
1-loop contribution which does not scale like the tree
approximation (section 2). Even if the renormalization is performed at
finite temperature such that the low energy constants (LECs) in the
effective action (in particular the pion decay constant) and
hence also the tree approximation become temperature-dependent the
remaining 1-loop contribution, which cannot be omitted, destroys
these simple scaling relations (section 3).

\section{1-loop at finite temperature}
Starting-point of our investigtion is the standard chiral lagrangian 
as given by Gasser and Leutwyler \cite{gl84}
\begin{eqnarray}
\label{lagrangian}
{\cal L}(U) &=& \frac{f^2}{2} 
\left[ \mbox{\boldmath$ \alpha$}_{\mu} \mbox{\boldmath$ \alpha$}_{\mu}
-2 m^2 u \right] \nonumber \\
&+& \ell_2 (\mbox{\boldmath$ \alpha$}_{\mu} \times \mbox{\boldmath$ \alpha$}_{\nu}
)^2 - (\ell_1+\ell_2) (\mbox{\boldmath$ \alpha$}_{\mu} 
\mbox{\boldmath$ \alpha$}_{\mu})^2 
+ \ell_4 m^2 (\mbox{\boldmath$ \alpha$}_{\mu} \mbox{\boldmath$
\alpha$}_{\mu}) u - (\ell_3 + \ell_4) m^4 u^2 \nonumber \\ 
&+& \cdots \nonumber \\
&=& \frac{f^2}{2} 
\mbox{\boldmath$ \alpha$}_{\mu} \mbox{\boldmath$ \alpha$}_{\mu}
-f^2 m^2 u 
+ \sum_i^4 \ell_i {\cal{L}}_i^{(4)} \, + \cdots 
\end{eqnarray}
with the abbreviations
\begin{eqnarray}
U^{\dagger} D_\mu U 
\equiv i\mbox{\boldmath$ \tau \cdot \alpha$}_{\mu}, \quad 
\frac{1}{4} tr(U+U^{\dagger}) \equiv u, \quad U  \in SU(2) \, . \nonumber
\end{eqnarray}
The first term in (\ref{lagrangian}) represents the N$\ell\sigma$ 
(nonlinear sigma) model which is of chiral order Ch${\cal{O}}(2)$ 
followed by the relevant Ch${\cal{O}}(4)$ terms.
For the time being the LECs take their standard values at scale 
$\mu=m_\rho$ in order to guarantee that
the lagrangian be leading order in the number of colors $N_C$. 
In the following we intend to use the same effective action in the
vacuum sector and in the soliton sector, the necessary modifications
will be discussed immediately. For the description of 1-loop effects
fluctuations {\boldmath$ \eta$}
\be \label{fluc}
U = \sqrt{U_0} \, \exp(i \mbox{\boldmath$ \tau$} \mbox{\boldmath$ \eta$}
/f ) \, \sqrt{U_0}
\ee
are introduced around the classical solution $U_0$
\begin{eqnarray} 
\label{hedgehog}
&U_0 = 1 \qquad \qquad \qquad & \mbox{vacuum sector} 
\nonumber \\
&U_0 = \exp( i \tau_a D_{ab} \hat r_b F(r) ) \qquad & \mbox{soliton sector}  
\end{eqnarray} 
which, in the soliton sector, is of the familiar hedgehog type
(rotational degrees of freedom are considered in the Wigner $D$-function,
translations are unimportant in this context). The different classical
solutions in the two sectors lead to decisive implications for the
proper 1-loop treatment: while in the vacuum sector 
perturbation theory (ChPT) applies as far as the external momenta are kept
small, the soliton sector's Casimir energy has to be evaluated via the
phase-shifts generated by the scattering equations for the fluctuations
which sum the 1-loop contribution to {\em all\/} chiral orders. This
complication is caused by the gradients of the chiral profile in (\ref{hedgehog})
which are of the order $m_{\rho}$ and not small.

For the same reason, in the soliton sector the higher chiral 
order terms in the lagrangian
(\ref{lagrangian}) are important  and may not
simply be neglected. The most elegant way to add higher chiral orders
to the effective action proceeds through the coupling of vector mesons
which leave the lagrangian to Ch${\cal{O}}(4)$ untouched. However,
unfortunately vector mesons come along with numerous technical
difficulties (many more degrees of freedom, induced components etc.)
whose proper treatment becomes forbiddingly complicated. The most simple
alternative is to use an {\em effective\/} 
LEC $\ell_2 = 1/4e^2$ in front of the Skyrme term.
The choice $e=4.25$ (instead of $e=7.24$)
simulates the higher chiral order terms generated by vector 
mesons \cite{mw96}. 
\begin{table}[h]
\begin{center} 
\begin{minipage}{10.0cm}
\baselineskip=12pt
{\begin{small}
Table 1. Tree contribution to the soliton mass and the baryon radius
for the Ch${\cal{O}}(4)$ lagrangian with LECs at scale $\mu = m_\rho$ 
($e=7.24$) and for the same lagrangian with an effective Skyrme
parameter $e=4.25$, compared with typical vector meson results
(see e.g. \cite{swhh89}).
\end{small}}
\end{minipage}
\begin{tabular}{|c|c|c|}
\hline
                   & soliton mass & baryon radius \\
                   &   $M [MeV]$  &  $<r^2>_B [fm^2]$ \\
\hline
ChO(4) lagrangian &   &   \\
(e=7.24)          & 940 & 0.09 \\
\hline
eff. Skyrme term  &   &   \\
 e=4.25           & 1630 & 0.24 \\
\hline
typical vector    &   &   \\
meson lagrangian  & $\simeq$ 1500-1600 & $\simeq$ 0.2-0.3\\
\hline
\end{tabular}
\end{center} 
\end{table} 
Tree contributions to soliton mass and radius of the
standard Ch${\cal{O}}(4)$ lagrangian 
with and without effective Skyrme prameter
are compared with typical vector
meson results in Table 1.
For $e=7.24$ the tree mass of $940 MeV$  
is reduced to $385 MeV$ if 1-loop corrections are included.
This soliton is just too small. With the effective Skyrme parameter
$e=4.25$ the tree values are of the typical magnitude obtained also
from vector meson models. For a more detailed justification of this choice
see ref. \cite{mw96}.

Observables other than the mass may be calculated
by coupling to the corresponding external fields.
External electromagnetic and axial fields couple through the covariant
derivative $D_\mu$ and the external scalar field, relevant for quark
condensate and sigma term, couples to the quark mass contained in the parameter
$m^2$. 
In general the external field
$j$ with strength $\varepsilon$ adds to the lagrangian in the form
\be \label{chilaex}
{\cal{L}}(\varepsilon) = {\cal{L}}(\ell_i, U) - 
\varepsilon j \cdot J(\ell_i, U) \; ,
\ee
where $J(\ell_i, U)$ denotes the corresponding current density. The
external field has to be chosen suitably so as to give the desired
quantity, for details see ref. \cite{mw96}.
Matrix elements of $j \cdot
J$ are then obtained as a derivative of the soliton mass (tree + 1
loop) in the presence of the external field with respect to its
strength. 

Later on all thermodynamical quantities of the vacuum and soliton
sector respectively, such as thermodynamical potentials,
entropy and so on are derived from the partition functions ($\beta=T^{-1}$)
\begin{eqnarray} 
\label{party}
\ell n Z _{vac} & = & \beta V \left[ f^2 m^2 + \frac{3}{2} g_0(m,T) \right] 
\nonumber \\
\ell n Z \quad & = & \ell n Z_{vac} - \beta M - \beta F_{cas} \, . 
\end{eqnarray} 
The vacuum partition function comprises the tree contribution $f^2 m^2$
of the lagrangian (\ref{lagrangian}) with $U_0=1$ and the familiar 1-loop
contribution for free massive pions $\frac{3}{2} g_0(m,T)$ (\ref{heat}).
Analogously, in the soliton sector there appears the soliton mass $M$
in tree approximation and the temperature-dependence  
resides for the time being in the free Casimir energy
$F_{cas}$ (1-loop contribution). The Casimir energy is generated 
by the fluctuations off the soliton background and
may be evaluated by means of the phase-shift formula \cite{m93}
\begin{eqnarray} 
F_{cas} & = & \frac{1}{\pi} \int_0^{\infty} dp
\delta^\prime (p) \left[ \frac{\omega}{2} + \frac{1}{\beta}
\ell n ( 1 - e^{- \beta \omega})  \right]   \\
& = & - \frac{1}{\pi} \int_0^{\infty} dp
\frac{p}{\omega} \delta (p) \left[ \frac{1}{2} + \frac{1}{e^{\beta
\omega}-1} \right] 
- \frac{\delta (0)}{\pi} \left[ \frac{m}{2} + \frac{1}{\beta}
\ell n ( 1 - e^{- \beta m})   \right] \, . \nonumber
\end{eqnarray}
This expression is ultraviolet divergent and requires renormalisation.
It should be noticed here that the divergence is located in the
temperature-independent part, the temperature-dependent part is
perfectly finite and does not introduce new infinities 
(see eq.(\ref{explicit}) below).
The divergencies are related to the high momentum behaviour of the
phase-shifts
\be
\delta (p)  \stackrel{p \to \infty}{\longrightarrow} a_0 p^3 + a_1 p +
\frac{a_2}{p} + \cdots \quad , 
\ee
(the denoted terms give rise to at least logarithmically divergent
expressions). The constants $a_0$, $a_1$ and $a_2$ are related to the
corresponding heat kernels and are known analytically for the
N$\ell\sigma$ model (first term in eq. (\ref{lagrangian}))
\begin{eqnarray}
\label{a}
a_0 &=& 0 \nonumber \\
a_1 &=& \frac{1}{4 \pi} \int d^3\!r \,\,\left[
2 \mbox{\boldmath$ \alpha$}_{\mu} \mbox{\boldmath$ \alpha$}_{\mu}
-3 m^2 (u-1) \right] \\
a_2 &=& \frac{1}{8 \pi} \int d^3\!r \,\,\left[
\frac{1}{3} (\mbox{\boldmath$ \alpha$}_{\mu} \mbox{\boldmath$ \alpha$}_{\mu}
)^2 + \frac{2}{3} (\mbox{\boldmath$ \alpha$}_{\mu} 
\mbox{\boldmath$ \alpha$}_{\nu})^2  \right. \nonumber \\
&& \left. \qquad \qquad \qquad \qquad \qquad
-2 m^2 (\mbox{\boldmath$ \alpha$}_{\mu} \mbox{\boldmath$
\alpha$}_{\mu}) (u-1) + \frac{3}{2} m^4 (u-1)^2 \right] \, . \nonumber 
\end{eqnarray}
Numerical values for the full model (\ref{lagrangian}) are 
$a_0=0.1 m_\pi^{-3}$, $a_1=3.6 m_\pi^{-1}$ and $a_2=15.2 m_\pi$. 
The strategy is now to subtract the troublesome terms in the
phase-shift integral and add them separately using dimensional
regularisation which involves a scale $\mu$ hidden in $\lambda,
G_0,G_1$ and $G_2$ 
\begin{eqnarray} 
F_{cas} & = & - \frac{1}{\pi} \int_0^{\infty} dp
\frac{p}{\omega} \left[ \delta (p) - a_0 p^3 - a_1 p - \frac{a_2}{p} \right]
\left[ \frac{1}{2} + \frac{1}{e^{\beta
\omega}-1} \right] \nonumber \\
&& - \frac{\delta (0)}{\pi} \left[ \frac{m}{2} + \frac{1}{\beta}
\ell n ( 1 - e^{- \beta m})   \right]    \\
&& + \lambda \left[ 3 \pi m^4 a_0 - 4 \pi m^2 a_1 + 8 \pi a_2 \right] 
-3 \pi a_0 G_0 - 2 \pi a_1 G_1 - 4 \pi a_2 G_2  \, \, .
\nonumber
\end{eqnarray}
The pole contributions ($d \to 4$) located in
\be \label{lambda}
\lambda = \frac{\mu^{d-4}}{16 \pi^2} \left[ \frac{1}{d-4} - \frac{1}{2}
(\Gamma'(1) + \ell n (4 \pi) + 1) \right]
\ee
may be renormalized
\be \label{exp}
3 \pi m^4 a_0 - 4 \pi m^2 a_1 + 8 \pi a_2 = \sum_i
\gamma_i \int d^3\!r\,\, {\cal{L}}_i^{(4)}  + \cdots
\ee
into the Ch${\cal{O}}(\geq 4)$ terms of the lagrangian. In Ch${\cal{O}}(4)$
the coefficients $\gamma_i$ \cite{gl84} are simple numerical factors
and the pole contributions are absorbed in the renormalized LECs
$\ell_i \to \ell_i + \gamma_i \lambda$ just as in standard ChPT. 
The remaining expression for the Casimir energy
\begin{eqnarray}
\label{Casimir} 
F_{cas} & = & - \frac{1}{\pi} \int_0^{\infty} dp
\frac{p}{\omega} \left[ \delta (p) - a_0 p^3 - a_1 p - \frac{a_2}{p} \right]
\left[ \frac{1}{2} + \frac{1}{e^{\beta
\omega}-1} \right] \\
&& - \frac{\delta (0)}{\pi} \left[ \frac{m}{2} + \frac{1}{\beta}
\ell n ( 1 - e^{- \beta m})   \right]    
-3 \pi a_0 G_0 - 2 \pi a_1 G_1 - 4 \pi a_2 G_2  \nonumber
\end{eqnarray}
is finite: it is the central formula of our formulation. The explicit
temperature and scale-dependence is contained in the contributions
\begin{eqnarray}
\label{G} 
G_0 & = & g_0(m,T) - \frac{m^4}{32 \pi ^2} (\frac{1}{6} + \ell n
\frac{m^2}{\mu^2}) \nonumber \\ 
G_1 & = & g_1(m,T) + \frac{m^2}{16 \pi ^2} \ell n \frac{m^2}{\mu^2}) \\ 
G_2 & = & g_2(m,T) - \frac{1}{16 \pi ^2} (1 + \ell n \frac{m^2}{\mu^2})
\, .
\nonumber
\end{eqnarray}

Let us postpone the discussion of the temperature-dependence 
together with the definition of the heat functions $g_\nu (m,T)$ for a moment.
The explicit scale dependence in (\ref{G}) should be compensated for by
the scale dependent LECs. This is actually the case also in the soliton
sector: at $T=0$ the LECs introduce a scale dependence to the soliton
in tree approimation. This scale dependence is compensated for by the
1-loop contribution such that the mass and
other baryon properties remain independent from the scale $\mu$ over a
wide region (Figs. 3.2 and 3.3 in ref. \cite{mw96}). Towards smaller
values then ($\mu \simeq 420 MeV$) the symmetric Ch${\cal{O}}(4)$
term $(\mbox{\boldmath$ \alpha$}_{\mu} \mbox{\boldmath$ \alpha$}_{\mu})^2$
in the lagrangian (\ref{lagrangian}) becomes too strong
and finally destroys the soliton. The renormalization scheme relies
on the premises of the existence of such a scale region with almost
constant baryon properties which therefore 
supports the reasonable choice for the
effective Skyrme parameter $e=4.25$. For the standard ChPT value
$e=7.24$ there appears a severe scale dependence. 
Typical tree and 1-loop values for
several observables (mass $M$, sigma term $\sigma$, axial coupling
$g_A$, isovector magnetic moment $\mu_V$ and electric polarizability
$\alpha$) are given in Table 2.
\begin{table}[h]
\begin{center} 
\begin{minipage}{13.5cm}
\baselineskip=12pt
{\begin{small}
Table 2. Tree and tree + 1-loop values for some typical
observables considered for the model with effective Skyrme parameter
$e=4.25$. For the axial coupling constant a
$1/N_C$ piece estimated from current algebra is added.
\end{small}}
\end{minipage}
\begin{tabular}{|c|c|c|c|c|}
\hline
& tree & tree + 1-loop & tree + 1-loop & experiment \\
& ${\cal{O}}(N_C)$ & ${\cal{O}}(N_C)+{\cal{O}}(1)$ & + CA correction &  \\
\hline
$M [MeV]$ & 1630 & 946 &  & 938 \\
\hline
$\sigma$ $[MeV]$ & 54 & 33 &    & 45 $\pm$ 7 \\
\hline
$g_A$ & 0.91 & 0.66 & 1.20 & 1.26 \\
\hline
$\mu^V$ & 1.62 & 2.24 &     & 2.35 \\
\hline
$\alpha$ $[10^{-4} fm^3]$ & 17.8 & 9.8 &  & 9.5 $\pm$ 5 \\
\hline
\end{tabular}
\end{center} 
\end{table} 
The 1-loop contributions generally improve the tree values towards
the experimental values except for the axial quantities. Because of the
Adler-Weissberger relation $g_A^2 = 1 + \, \cdots \, $ a
large $1/N_C$ contribution to $g_A$ is expected \cite{mw96}.
This unpleasent feature caused by the algebra of the axial currents
which mixes different $N_C$ orders is common to all models relying
on the $1/N_C$ expansion.

The temperature-dependence of the Casimir energy
(\ref{Casimir}) is made explicit by writing
\begin{eqnarray}
\label{explicit} 
F_{cas} (T) & = & F_{cas} (0) +
\frac{1}{\pi \beta} \int_0^{\infty} dp
\delta^\prime (p)  \ell n ( 1 - e^{- \beta \omega}) \, .  
\end{eqnarray}
As already mentioned the temperature-dependent part is finite 
and requires no extra
renormalization. The important contributions come from the terms
proportional to the heat functions contained in the terms
$G_0, G_1, G_2$ (last column in eq. (\ref{Casimir}))
\begin{eqnarray}
\label{heat}
g_0(m,T) &=& \frac{2}{3(2\pi)^3} \int d^3\!p \,\,
\frac{p^2}{\omega} \frac{1}{e^{\beta \omega}-1} \,\,\,
\stackrel{T>>m}{=} \quad 
\frac{\pi^2}{45} T^4 - \frac{m^2 T^2}{12} + \frac{m^3 T}{6\pi}
+ \cdots \nonumber \\
g_1(m,T) &=& \frac{1}{(2\pi)^3} \int d^3\!p \,\,
\frac{1}{\omega} \frac{1}{e^{\beta \omega}-1}
\quad \, \, \, \stackrel{T>>m}{=} \quad \frac{T^2}{12} - \frac{m T}{4\pi}
+ \cdots \\
g_2(m,T) &=& \frac{1}{2(2\pi)^3} \int d^3\!p \,\,
\frac{1}{p^2 \omega} \frac{1}{e^{\beta \omega}-1}
\, \stackrel{T>>m}{=} \quad \frac{T}{8\pi m}
+ \cdots \, , \nonumber 
\end{eqnarray}
which are of Ch${\cal{O}}(4)$, Ch${\cal{O}}(2)$ and Ch${\cal{O}}(0)$
respectively. Although we could have integrated (\ref{explicit}) directly,
we use (\ref{Casimir},\ref{G}) with the heat functions (\ref{heat}) 
to make the connection to the renormalization at finite temperature
(next section) more transparent.

The temperature-independent tree and the temperature-dependent 1-loop 
contributions to the free and internal soliton energies
$F_{sol}$ and $U_{sol}$ 
\begin{eqnarray}
F_{sol} & \equiv & - \frac{\ell n Z/Z_{vac}}{\beta} \quad = \, M + F_{cas}
\nonumber \\ 
U_{sol} & \equiv & - \frac{\partial \ell n Z/Z_{vac}}{\partial \beta} \, = \, 
M + F_{cas}-\frac{\partial F_{cas}}{\partial T} \, ,
\end{eqnarray}
as well as to the axial coupling constant $g_A$ are shown
in Figs. 3 and 4 (dashed lines). Both energies $F_{sol}$ and
$U_{sol}$ remain almost constant over the
low temperature region. Towards the critical temperature  
$F_{sol}$ decreases and $U_{sol}$ 
increases as expected, but there we cannot trust the 1-loop
approximation (next section). The different behaviour of free and
internal energies is not surprising: this is observed already
for massless bosons in the vacuum (compare also (\ref{limit})). 
The axial coupling $g_A$ is more sensitive:
it decreases with increasing temperature already at relatively 
low temperatures. 

Apparently there is no simple scaling law. Even in the chiral limit
$m \to 0$ where the Casimir energy remains perfectly well
defined (because of $\delta^{\prime} (0)=0$ for $m=0$) the 
temperature-dependence by expection of (\ref{explicit}) is non-trivial. 
Therefore in the following section we perform the renormalization at
finite temperature in order to obtain a temperature-dependent soliton
(tree approximation). Because the scaling (\ref{scaling}) of the
tree approximation is simple one might expect a similar behaviour
for the total temperature-dependent contribution. This expection
will proof wrong.

\section{Renormalisation group at finite temperature}

Because the temperature-dependent contribution to the Casimir energy is
finite there is of course some ambiguity in setting up the
renormalization scheme.
Among the three relevant terms in the Casimir
energy (\ref{Casimir}), the first one proportional to $a_0$ is of
higher chiral order (at least Ch${\cal{O}}(6)$) and numerically
irrelevant. For the remaining terms proportional to $a_1$ and $a_2$
there are several possibilities:
\begin{itemize}
\item[(i)] no renormalization: both terms are kept in the 1-loop
contribution, the renormalized LECs take their values at $T=0$. This
was the choice made in section 2.
\item[(ii)] minimal renormalization: the $a_1$-term is renormalized into
the N$\ell\sigma$ model (first term in (\ref{lagrangian})) and renders
the parameters $f$ and $m$ temperature-dependent. The $a_2$-term is
kept in the 1-loop contribution, the Ch${\cal{O}}(4)$ LECs 
$\ell_i (\mu)$ are untouched and remain at their $T=0$ values. The
resulting temperature-dependence of $f$ and $m$ will proof to be in
accordance with the expectations of ChPT.
\item[(iii)] maximal renormalization: both, 
the $a_1$ and the $a_2$-term are renormalized into the N$\ell\sigma$
model and the Ch${\cal{O}}(4)$ piece of the lagrangian, respectively.
Because in the original lagrangian there are no other terms this
corresponds to a maximal renormalization of the temperature-dependent
1-loop. Now also the $\ell_i$s become temperature-dependent 
$$ \ell_i (\mu,T) = \ell_i (\mu) + \frac{\gamma_i}{2} g_2(m,T) $$
through the heat function $g_2(m,T)$. Unfortunately this function
diverges in the chiral limit $m \to 0$ (compare (\ref{heat})) and hence
also the renormalized $\ell_i$s, leading to an ill-defined lagrangian.
Thus, in this renormalization scheme the total finite
temperature-dependent contribution (\ref{explicit}) is artificially
split into infinite tree and 1-loop contributions for
vanishing pion mass, which is very unpleasant.
\item[(iv)] maximal renormalization with temperature-dependent scale
$\mu$: by choosing a suitable temperature-dependent scale
$\ell n \mu^2/m_\rho^2 = 16 \pi^2 g_2$ the LECs of Ch${\cal{O}}(4)$
may be reset to their original values at $\mu = m_\rho $
$$ \ell_i (\mu,T) = \ell_i (m_\rho) - \frac{\gamma_i}{32 \pi^2}
\ell n \frac{\mu^2}{m_\rho^2} + \frac{\gamma_i}{2} g_2(m,T)
= \ell_i (m_\rho) \, .$$
The $\ell_i$s become now temperature-independent again but the scale
transformation shows up in the 1-loop contribution
through the functions $G_\nu$ (\ref{G}) producing exactly the terms
proportional to $g_2(m,T)$ which were renormalized away in 
(iii). This scheme becomes essentially identical to the minimal
renormalization (ii) which amounts to the trivial statement that by the
scale choice finite contributions are shifted from tree to 1-loop and
vice versa.
\end{itemize} 
There is another unpleasant feature of the maximal renormalization
scheme: because the coefficient $\ell_1+\ell_2$ in front of the symmetric
Ch${\cal{O}}(4)$ term grows rapidly with increasing temperature
the soliton becomes unstable already at relatively low temperatures
below $100 MeV$. This unphysical phenomenon is due to the lagrangian
(\ref{lagrangian}) restricted to Ch${\cal{O}}(\leq 4)$; with explicit 
sigma mesons there appears no such difficulty.

For our further discussion we will therefore choose the minimal
renormalization scheme (ii) with temperature-independent $\ell_i$s.
The corresponding renormalisation group (RG) equations
\cite{mnu84,finy86,p92} for the parameters $f$ and $m$ 
of the N$\ell\sigma$ model are written in differential form
\begin{eqnarray}
\label{rge}
d f^2 & = & \, \quad -2  \frac{\partial g_1}{\partial T} d T
\, = \, -2 (dg_1 + g_2 dm^2) \nonumber \\
d (f^2 m^2) & = & -\frac{3}{2} m^2  \frac{\partial g_1}{\partial T} d T
\, = \, -\frac{3}{2} m^2 (dg_1 + g_2 dm^2) \, . 
\end{eqnarray}
These equations, augmented by the corresponding expression for the
vacuum partition function (\ref{party})
\be
\label{rge0}
d (\frac{\ell n Z_{vac}}{\beta V}) = \frac{3}{2} 
\frac{\partial g_0}{\partial T} d T
\, = \, d(f^2 m^2) + \frac{3}{2} d(g_0 + m^2 g_1) + 
\frac{3}{2} m^2 g_2 dm^2 \, , 
\ee
are solved numerically. In order to obtain an analytical solution
which can be compared to ChPT the terms proportional to $dm^2$ may be
neglected. Because $dm^2$ is at least Ch${\cal{O}}(4)$ 
(see eq.(\ref{fpi}) below)
these terms contribute to the action only at Ch${\cal{O}}(\geq 6)$
where many other contributions are also omitted. 
On the other hand
by solving the RG equations in this way we sum up a class of diagrams
to {\em all\/} chiral orders: these are just
the chain, daisy and super-daisy graphs
\cite{ny96} as will be noticed immediately. 
\begin{figure}[htbp]
\vspace*{5.5cm}
\begin{center} 
\begin{minipage}{10.0cm}
\baselineskip=12pt
{\begin{small}
Fig. 1. Temperature dependence of the pion decay constant $f$ and the pion mass
$m$ as derived from the RG equation (full lines).
In the chiral limit (dashed lines) the pion decay
constant vanishes and the pion mass diverges at the critical
temperature. All quantities in this and the following figures are
plotted relative to their values at $T=0$.
\end{small}}
\end{minipage}
\end{center}
\end{figure}
The solution, where quantities
evaluated at $T=0$ are marked by the subscript zero, is
\begin{eqnarray}
\label{fpi}
f^2 &=& f_0^2 - 2 g_1(m,T)  \nonumber \\     
m^2 &=& m_0^2 \left[1 - \frac{2 g_1(m,T)}{f_0^2} \right]^{-\frac{1}{4}} \\
\frac{\ell n Z_{vac}}{\beta V} &=& f^2 m^2 + \frac{3}{2} \left[
g_0(m,T) + m^2 g_1(m,T) \right] \, . \nonumber
\end{eqnarray}
These expressions, exact in the chiral limit, were checked against 
the numerical solution of the RG equations (\ref{rge},\ref{rge0}): there are only
marginal deviations at higher temperatures.
The parameters $f$ and $m$ are plotted in Fig. 1. 
If (\ref{fpi}) is systematically expanded to Ch${\cal{O}}(8)$ 
\begin{eqnarray}
\frac{\ell n Z_{vac}}{\beta V} & = & f_0^2 m_0^2 + \frac{3}{2} g_0(m_0,T)
+ \frac{3 m_0^2 g_1^2(m_0,T)}{8 f_0^2} \\
&& \qquad \qquad \quad
+\frac{m_0^2 g_1^2(m_0,T)}{16 f_0^2} \left[ 5 g_1(m_0,T) - 
3 m_0^2 g_2(m_0,T) \right] + \cdots \, , \nonumber
\end{eqnarray}
the ChPT result of \cite{gl89} reproduced except for one
diagram of Ch${\cal{O}}(8)$ which is not of the chain or daisy type.
Although the RG improvement is meant to extend the 1-loop results
to higher temperatures and although the relations
(\ref{fpi}) describe a second order chiral phase transition (subsequent
subsection) we have to be cautious: because the model does not explicitely
include heavier
mesons like vector mesons and especially the scalar meson which plays a
crucial role in chiral symmetry restoration the results may not be 
trusted close to the phase transition, e.g. 
the critical temperature turns out much too high here.

\subsection{Vacuum sector}

All thermodynamical properties of the vacuum are derived from the
partition function $Z_{vac}$ (\ref{fpi}). Examples are the free and
internal vacuum energy densities
\begin{eqnarray}
\label{pot}
F_{vac}/V & \equiv & - \frac{\ell n Z_{vac}}{\beta V} \quad = \, 
- \frac{3}{2} \left[ g_0(m,T) + m^2 g_1(m,T) \right] -f^2 m^2
\nonumber \\ 
U_{vac}/V & \equiv & - \frac{\partial \ell n Z_{vac}}{V \partial \beta}
\, = \, + \frac{3}{2} \left[ 3g_0(m,T) + m^2 g_1(m,T) \right] -f^2 m^2
\end{eqnarray}
and  the quark condensate. 
\begin{figure}[h]
\vspace*{5.5cm}
\begin{center} 
\begin{minipage}{10.0cm}
\baselineskip=12pt
{\begin{small}
Fig. 2. Temperature dependence of the
quark condensate as derived from the RG equation (full line).
In the chiral limit (dashed line) the quark condensate
vanishes at the critical temperature.
For comparison the 1-, 2- and 3-loop ChPT results are also depicted
(dotted lines).
\end{small}}
\end{minipage}
\end{center}
\end{figure}
The quark condensate is obtained as a derivative
with respect to the quark mass $m_q \sim m_0^2$:
\begin{eqnarray}
\label{cond}
<\bar q q> & \equiv & -\frac{\partial}{\partial m_q} 
(\frac{\ell n Z_{vac}}{\beta V}) =
\frac{<\bar q q>_0}{f_0^2}
\frac{\partial}{\partial m_0^2} 
(\frac{\ell n Z_{vac}}{\beta V}) \nonumber \\ 
<\bar q q> &=& <\bar q q> _0 \frac{f^2 m^2}{f_0^2 m_0^2} \,
= \, <\bar q q>_0 \left[1 - \frac{2 g_1(m,T)}{f_0^2} \right]^{\frac{3}{4}} 
\, . \nonumber
\end{eqnarray}
The relation $<\bar q q> \sim f^2 m^2$, which should hold for a
consistent RG scheme at finite temperature,
follows here from the
definition after some non-trivial algebra using (\ref{fpi}). The quark
condensate is compared with 1-, 2- and 3-loop ChPT results \cite{gl89}
in Fig. 2. The pion mass and the quark condensate scale with the pion
decay constant like
\be 
m \sim f^{-1/4} \, , \qquad <\bar q q> \sim f^{3/2}
\ee
which, although rigorous at low temperatures, is not meant to remain
valid close to the phase transition for the reasons already discussed. 

Nevertheless we briefly comment on the chiral limit $m_0 \to 0$
where the expressions (\ref{fpi},\ref{pot},\ref{cond}) simplify
(dashed lines in Figs.1 and 2) and on the chiral phase transition: 
\begin{eqnarray}
\label{limit}
f^2  \stackrel{m_0 \to 0}{\longrightarrow}  
f_0^2 \left[ 1 - \frac{T^2}{6 f_0^2} \right] \qquad \qquad \qquad    
&& m^2  \stackrel{m_0 \to 0}{\longrightarrow}  
m_0^2 \left[ 1 - \frac{T^2}{6 f_0^2} \right]^{-\frac{1}{4}}  \nonumber  \\
F_{vac}/V  \stackrel{m_0 \to 0}{\longrightarrow}   
- \frac{\pi^2}{30} T^4 \qquad \qquad \qquad
&& U_{vac}/V \stackrel{m_0 \to 0}{\longrightarrow}   
+ \frac{\pi^2}{10} T^4  \\
<\bar q q>  \stackrel{m_0 \to 0}{\longrightarrow}  
<\bar q q>_0 \left[ 1 - \frac{T^2}{6 f_0^2} \right]^{\frac{3}{4}} \ . 
&&  \nonumber
\end{eqnarray}
Pion decay constant $f$, quark
condensate $<\bar q q>$ and the gluon condensate $f^4/8 (\ell_1 +
\ell_2)$ become
zero at the same critical temperature $T_c = \sqrt{6} f_0$. The
diverging pion mass and the too large critical temperature are
artifacts of the missing heavier mesons in particular the sigma meson.
Critical exponents according to the standard definitions
(specific heat $C \equiv - T d^2F_{vac}/dT^2$)
\begin{eqnarray}
&& C \stackrel{m_q \to 0}{\sim} \vert T-T_c \vert^{\alpha} \nonumber \\
&& <\bar q q>  
\stackrel{m_q \to 0}{\sim} \vert T-T_c \vert^{\beta} \nonumber \\
&& \frac{\partial <\bar q q>}{\partial m_q}  
\stackrel{m_q \to 0}{\sim} \vert T-T_c \vert^{-\gamma}  \\
&& <\bar q q>  
\stackrel{T=T_c}{\sim} m_q^{1/\delta} \nonumber
\end{eqnarray}
are readily read off: $\alpha=0, \beta=\frac{3}{4}, \gamma=\frac{3}{8}, 
\delta=3$. At least the coefficients $\alpha$, $\beta$ and $\delta$ 
agree with the leading terms of an $\varepsilon$ - expansion
\cite{zj89} ($\gamma=\frac{3}{8}$ instead of $\gamma=\frac{1}{2}$ 
suffers from the approximation leading to (\ref{fpi})).
However they deviate substantially from those of the O(4) Heisenberg
magnet obtained in the {\em linear \/} sigma model \cite{w92,rw93}.
A smooth connection of the low temperature behaviour of all these
quantities as discussed here with the linear sigma model results at 
the critical temperature $T_c$ is highly desirable. A simple Pad\'e
approximation \cite{bk96,jk96} which connects the two temperature regions
can not be the solution to this problem.

\subsection{Soliton sector}

Via the temperature-dependence of $f$ and $m$ the soliton in tree
\begin{figure}[htbp]
\vspace*{5.5cm}
\begin{center} 
\begin{minipage}{10.0cm}
\baselineskip=12pt
{\begin{small}
Fig. 3. Temperature dependence of the free soliton energy $F_{sol}$ 
and the internal soliton energy $U_{sol}$.
The 1-loop calculations
(dashed) are compared to the RG results (solid).
\end{small}}
\end{minipage}
\end{center}
\end{figure}
approximation is now itself temperature-dependent and so are the
fluctuations and the phase-shifts.
In the end $\ell n Z/Z_{vac}$ (\ref{party}) is given by 
the temperature-dependent soliton mass plus
the Casimir energy (\ref{Casimir}), but now
without the term $2 \pi a_1 g_1$ which is already taken care of by the
RG equation.
The Casimir energy must not be forgotten: it destroys the
scaling behaviour of the tree approximation and brings the results back 
close to those of the 1-loop calculation (section 2).
The results for
the free and internal soliton energies as well as the axial coupling
constant and the pion-nucleon coupling are shown in Figs. 3 and 4.
The temperature-dependence of the nucleon mass turns out to be modest
(compared to that of the pion decay constant $f$). 
\begin{figure}[h]
\vspace*{5.5cm}
\begin{center} 
\begin{minipage}{10.0cm}
\baselineskip=12pt
{\begin{small}
Fig. 4. Temperature dependence 
of the axial coupling
$g_A$ and the pion-nucleon coupling $g_{\pi NN}$. The 1-loop calculations
(dashed) are compared to the RG results (solid).
\end{small}}
\end{minipage}
\end{center}
%\end{figure}
%\begin{figure}[h]
\vspace*{5.5cm}
\begin{center} 
\begin{minipage}{10.0cm}
\baselineskip=12pt
{\begin{small}
Fig. 5. Temperature dependence 
of the $\sigma$-term.
The 1-loop calculation
(dashed) is compared to the RG result (solid).
\end{small}}
\end{minipage}
\end{center}
\end{figure}
The coupling
constants tend to decrease with increasing temperature.
There are other quantities like the $\sigma$-term and the electric
polarizability which show a much more pronounced temperature-dependence. 
The $\sigma$-term plotted in Fig. 5 melts very
quickly with increasing temperature.
The RG result starts
to deviate from the 1-loop at relatively low temperatures,  
obviously higher loops are much
more important for this quantity as compared to the soliton energies
and coupling constants already discussed. For the electric
polarizability (not plotted) there is the opposite finding: 
it increases rapidly with increasing temperature.
However it seems questionable whether these
observables can be measured in a hot environment.

\section{Conclusion}

The chiral soliton model is most suitable for the study of the nucleon in
the heat bath of hot pions because it contains both 
ingredients which are necessary: the pions and the nucleon as chiral soliton in one and the
same model. The temperature-dependence of baryon properties enters via
the 1-loop contribution. For the most important quantities determining
the behaviour of hot nucleonic matter this leads to an almost constant
mass and to a modestly decreasing pion-nucleon coupling in the low
temperature regime.

In order to implement the temperature-dependence already on the tree
level (temperature-dependent soliton) the finite
temperature RG equations are studied. These comprise and extend the well
established ChPT results for the temperature-dependent pion
decay constant, pion mass and chiral quark condensate. There is no need
to lend these quntities from other models. Via these temperature-dependent 
parameters the soliton is now itself temperature-dependent,
however the remaining 1-loop contribution cannot be disregarded. If
this contribution is taken properly into account, the
nucleon mass and the axial coupling are again back close to the
na\"{\i}ve 1-loop result. For other quantities, as e.g. the sigma term and
the electric polarizability, which show a much more pronounced
temperature-dependence, the RG treatment deviates from the na\"{\i}ve
1-loop result with increasing temperature indicating the importance of
higher loop effects. Unfortunately these quantities are hardly
experimentally accessible in a hot environment.

No simple scaling law was found in this investigation. The scaling
hypothesis relies on the tree approximation and the 1-loop contribution
destroys this behaviour. If we allow for temperature-dependent
Ch${\cal{O}}(4)$ LECs according to the maximal renormalization scheme,
in particular for a temperature-dependent Skyrme parameter $e$, the
scaling behaviour is improved but still far from being
satisfactory. This is caused by the higher chiral order terms 
contained in the 1-loop which are important in the soliton sector.

The RG improvement presented here extends the rigorous 
ChPT results to higher temperatures and
leads to a second order phase transition.
Nevertheless, the validity of this investigation remains limited to the low
temperature region because of serious short-comings of the underlying model.
To extend the procedure towards the phase
transition a RG treatment which includes heavier mesons, in particular
the sigma meson, has to be developed. At present, there is not even a RG scheme
which connects the ChPT results in the low temperature region smoothly
with the O(4) symmetry expected in the linear sigma model at the
critical temperature. More than that, quark degrees of freedom which
cannot be treated within the framework of effective lagrangians may
certainly become important close to the phase transition \cite{kk95}.

\vspace*{1.0cm}

I have benefited from numerous discussions with G. Holzwarth, H. Geyer
and S. Marculescu.

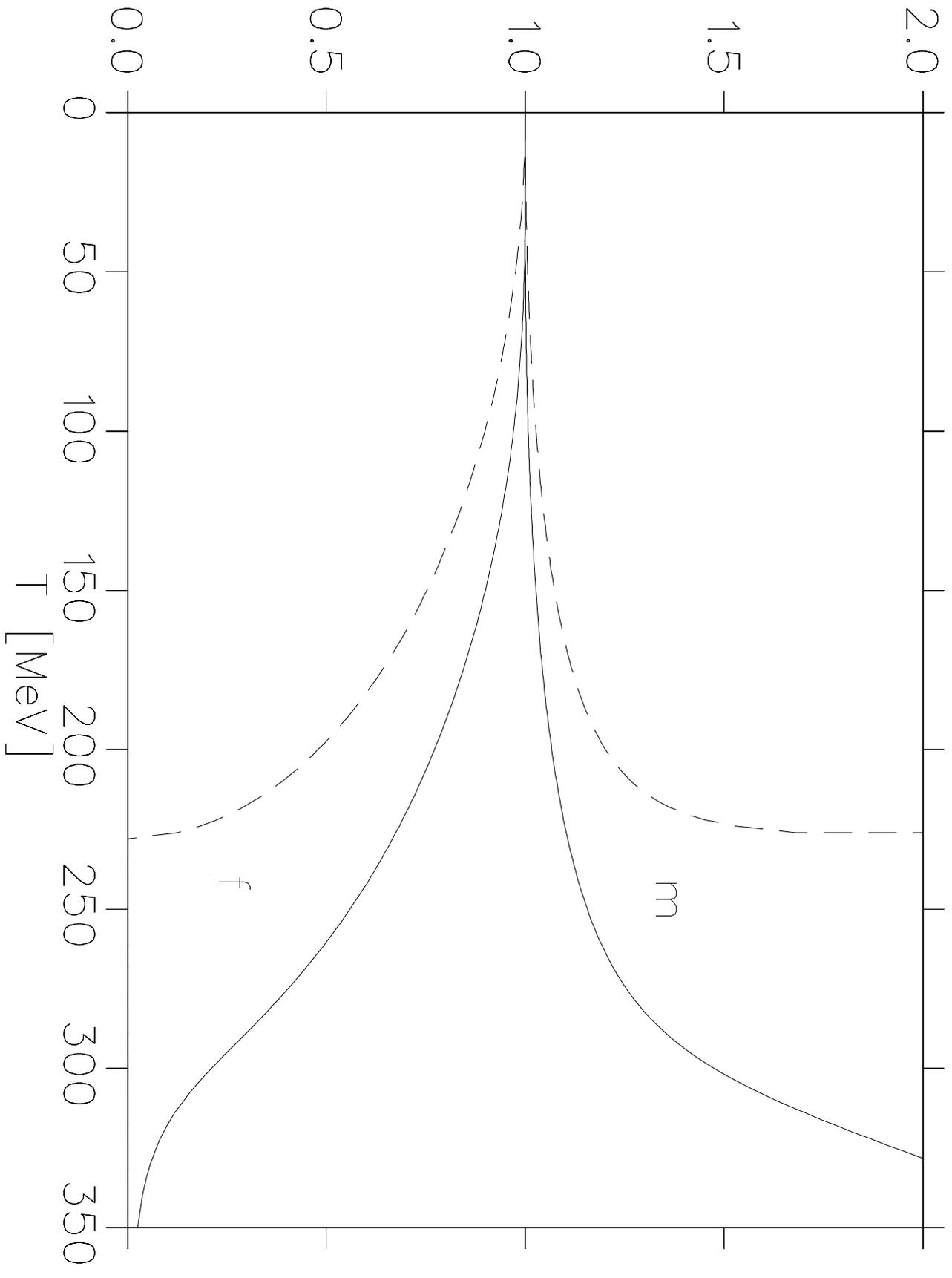
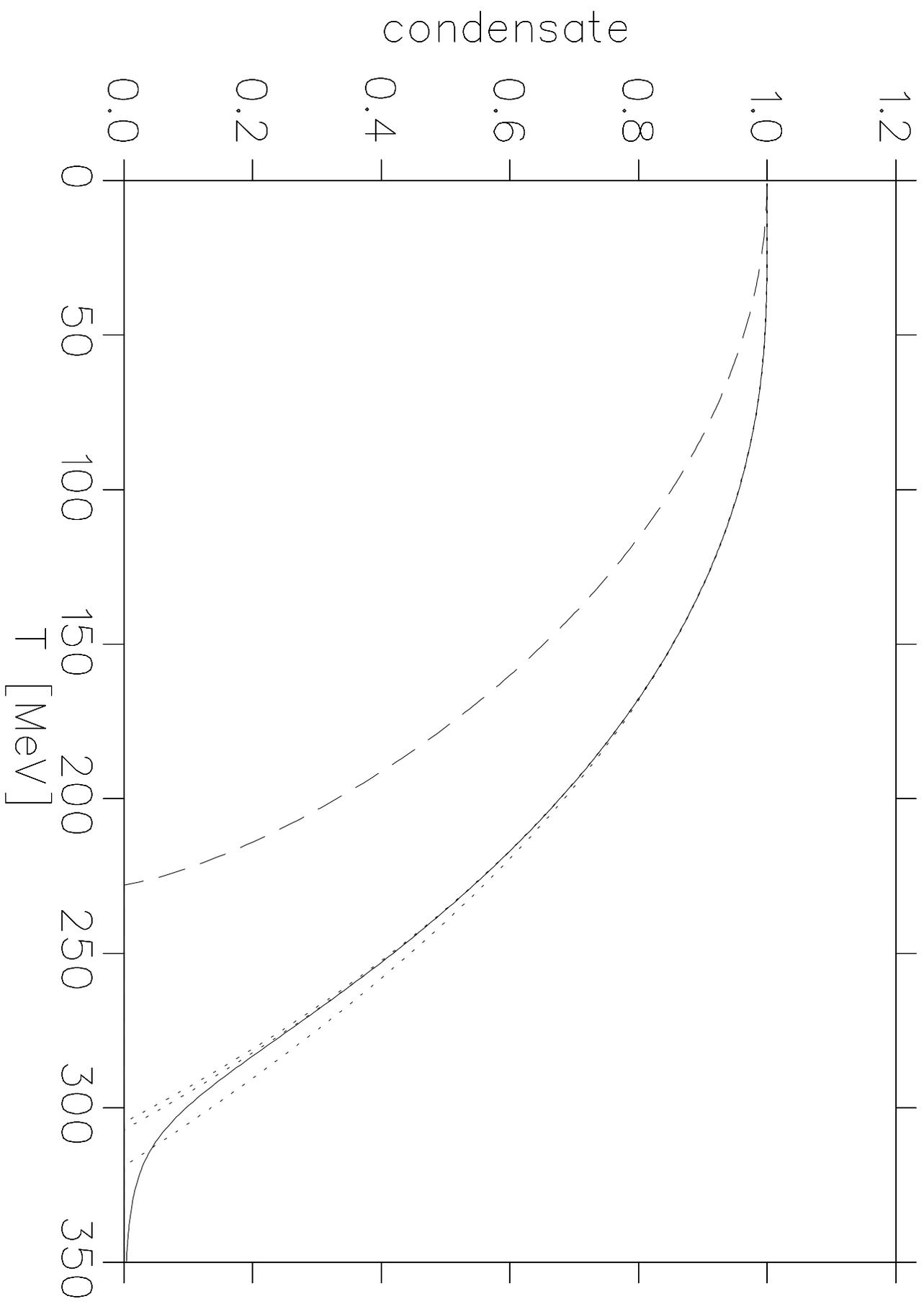
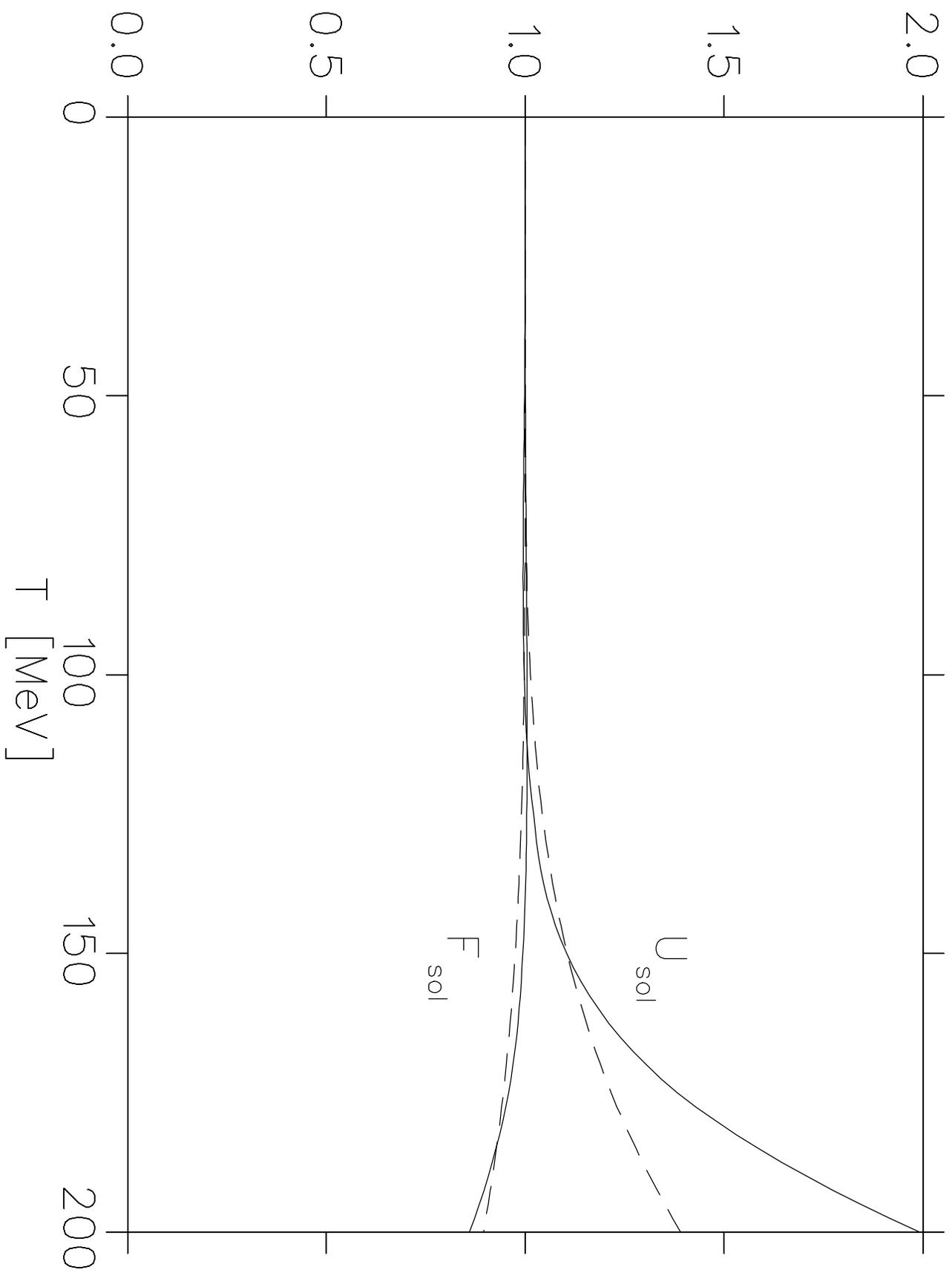
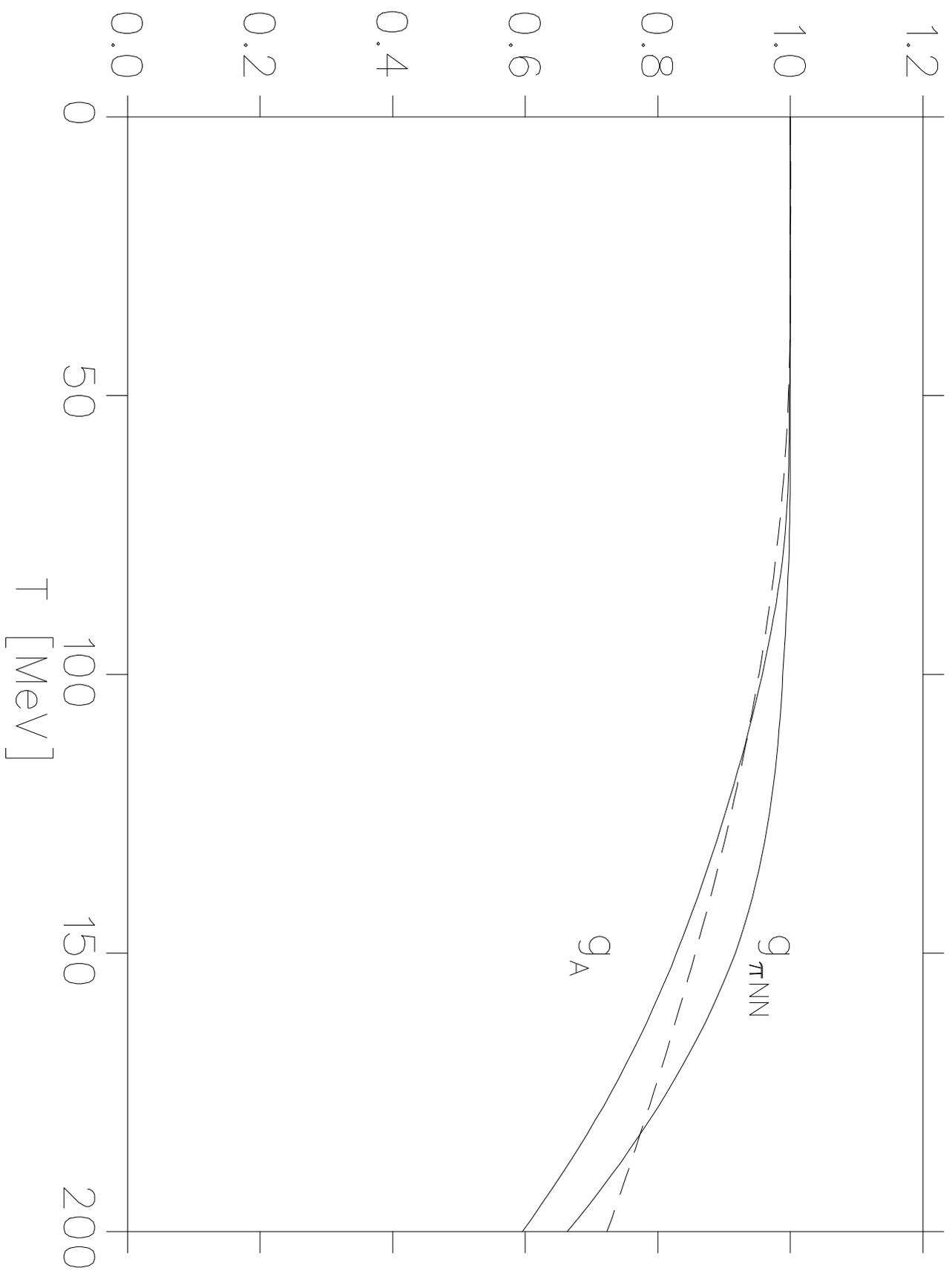
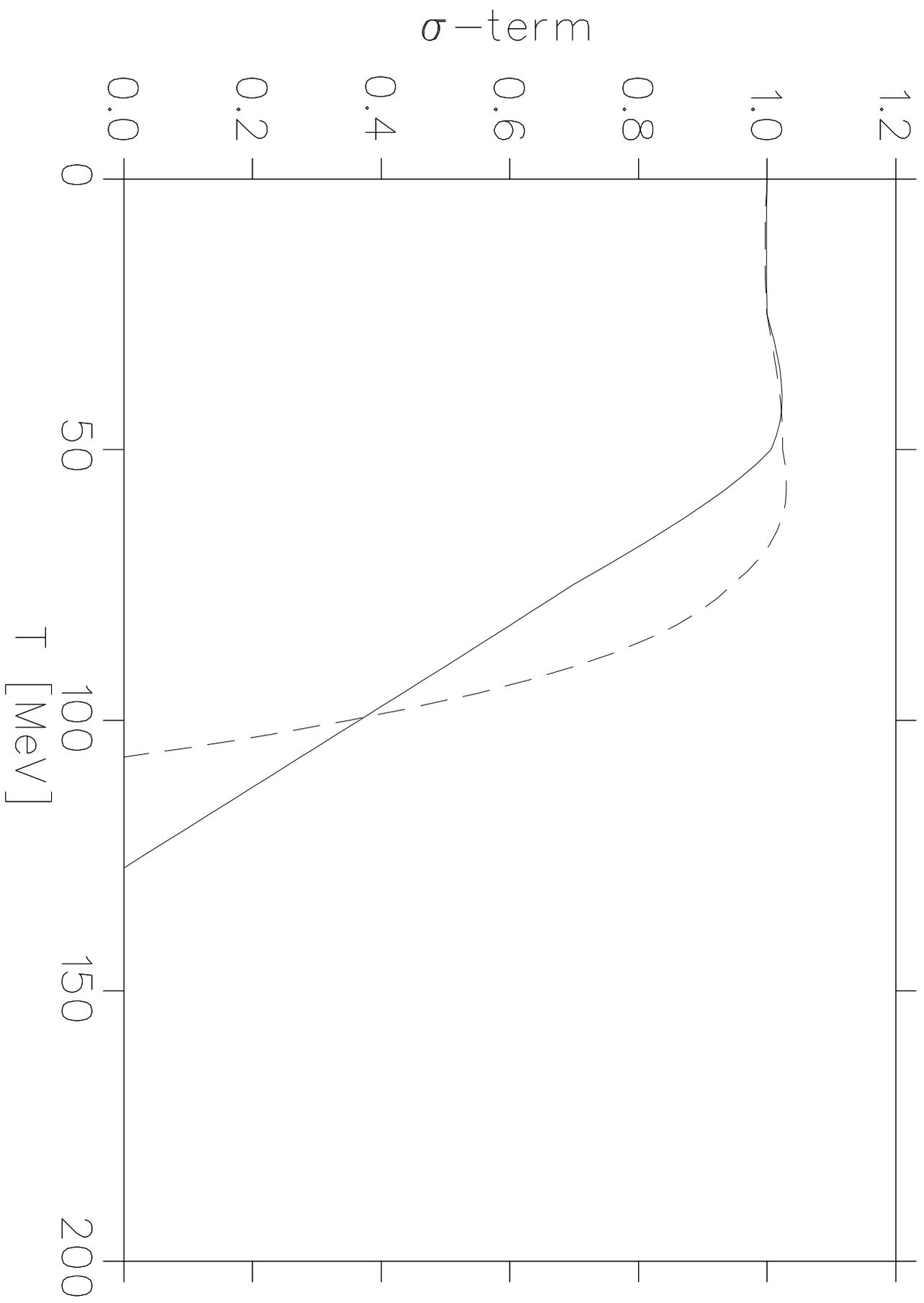
\end{document}